\documentstyle[12pt,preprint,prc,aps,psfig,floats]{revtex}
\input psfig
\def\lsim{\mathrel{\rlap{\lower4pt\hbox{\hskip1pt$\sim$}}
    \raise1pt\hbox{$<$}}}         %less than or approx. symbol
\def\gsim{\mathrel{\rlap{\lower4pt\hbox{\hskip1pt$\sim$}}
    \raise1pt\hbox{$>$}}}         %greater than or approx. symbol
\begin{document}
\draft

\title{How many sigmas is the solar neutrino effect?}
\author{{John N. Bahcall}\thanks{jnb@ias.edu}} 
\address{School of Natural Sciences, 
Institute for Advanced Study, Princeton, NJ 08540}
\maketitle
\begin{abstract}
The minimal standard electroweak model can be tested by allowing all
the solar neutrino fluxes, with undistorted energy spectra, to be free
parameters in fitting the measured solar neutrino event rates, subject
only to the condition that the total observed luminosity of the sun is
produced by nuclear fusion.  The rates of the five experiments prior
to SNO (chlorine, Kamiokande, SAGE, GALLEX, Super-Kamiokande) cannot
be fit by an arbitrary choice of undistorted neutrino fluxes at the
level of 2.5$\sigma$ (formally 99\% C.L.). Considering just SNO and
Super-Kamiokande, the discrepancy is at the 3.3$\sigma$
level($10^{-3}$ C.L.). If all six experiments are fit simultaneously,
the formal discrepancy increases to $4 \sigma$ ($7\times10^{-5}$
C.L.). If the relative scaling in temperature of the nuclear reactions
that produce $^7$Be and $^8$B neutrinos is taken into account, the
formal discrepancy is at the $7.4\sigma$ level.
\end{abstract}


\section{Introduction}

The Sudbury Neutrino Observatory (SNO) has reported an epochal
measurement~\cite{sno} of the rate of charged current interactions in
deuterium due to $^8$B solar neutrinos. The precise Super-Kamiokande
measurement~\cite{superk} of neutrino-electron scattering (charged
plus some neutral current sensitivity) by $^8$B neutrinos reveals a
total neutrino flux that is about $3.3 \sigma$~\cite{sno} larger than
the $\nu_e$ flux measured by SNO.  The combined SNO and
Super-Kamiokande result seems to have convinced most physicists that
neutrino oscillations are occuring in the solar neutrino domain.

Why is a single $3\sigma$ result so convincing? We all know of
 examples in particle and nuclear physics where three sigma results
have not been verified.  The purpose of this paper is quantify the
role of the additional information on solar neutrino event rates that,
when taken together with 
the SNO/Super-Kamiokande result, makes the inference
of neutrino oscillations so compelling.

I will not discuss the precise helioseismological verification, better
than $0.1$\% r.m.s. throughout the sun, of the sound speeds predicted
by the standard solar model~\cite{bp2000} (hereafter, BP2000).  I will
also not discuss non-quantifiable effects such as the manifestly great
care with which both the SNO and Super-Kamiokande experiments were
performed.  The excellent agreement between the standard solar model
predictions and the helioseismological measurements and the rigourous
calibrations of the SNO and Super-Kamiokande experiments are
undoubtedly important factors in convincing many in the physics
community that solar neutrino oscillations occur, but I will focus
here only on the measurements of solar neutrino event rates in
different detectors.

Suppose we allow the solar neutrino fluxes to have arbitrary
(positive) amplitudes, subject only to the conditions that the fusion
energy associated with these fluxes equals the precisely measured
solar luminosity and that the energy spectra are undistorted by
neutrino oscillations.  If the minimal standard electroweak model is
valid (no neutrino oscillations occur), solar neutrino energy spectra
differ from their laboratory shapes by of order one part in $10^5$ for
beta-decays (like $^8$B or $^{13}$N decay) and less than one part in
$10^2$ for the $p-p$ reaction(one part in $10^3$ for the $hep$
reaction)~\cite{bahcall94}.  Suppose we ignore all other information
about the sun, including helioseismology.  How well can we then fit
the observed set of solar neutrino event rates in different
experiments?

I report here on a simultaneous fit with arbitrary neutrino fluxes to
all the available neutrino event rates, chlorine~\cite{chlorine},
Kamiokande~\cite{kamiokande}, SAGE~\cite{sage}, GALLEX
+GNO~\cite{gallexgno}, Super-Kamiokande~\cite{superk}, and
SNO~\cite{sno}.  We shall see that including all of the available
experiments (not just SNO and Super-Kamiokande) increases by an order
of magnitude the stringency of the formal C.L. by which one can
conclude that neutrino oscillations are required (a $4\sigma$ effect
for all six experiments).  If the temperature scaling of the nuclear
reactions giving rise to $^7$Be and $^8$B neutrinos is imposed as an
additional condition on the fitting procedure, then the no-oscillation
hypothesis is rejected at the $7.4\sigma$ level (compared to
$6.9\sigma$ in the pre-SNO era).  

In the present paper, I also provide a physical explanation of why the
$\chi^2_{\rm min}$ method leads to the unphysical requirement that
some solar neutrino fluxes($^7$Be, $^{13}$N, and $^{15}$O) be
completely absent while the $p-p$ neutrino flux is enhanced over the
standard solar model prediction.

The calculations described in this paper utilize an improved
formulation of the solar luminosity constraint on neutrino
fluxes~\cite{luminositypaper}.

There have been a number of pre-SNO investigations of the failure of
free-flux, no-oscillation fits to solar neutrino data. The first such
study stressed as early as 1990~\cite{bethe90} the apparent
incompatability of the chlorine and Kamiokande experiments, if new
physics did not affect the shape of the solar neutrino energy
spectra. The seminal studies in the mid 1990s by Hata, Bludman, and
Langacker~\cite{hata94}, Parke~\cite{parke95}, and Heeger and
Robinson~\cite{heeger96}, and related discussions~\cite{spiro90},
helped convince many physicists of the necessity of solar neutrino
oscillations. The inadequacy of free-flux fits was reinforced by the
most recent pre-SNO studies~\cite{bahcallkrastev96,bks98}.

The principal results of this paper are presented in
Table~\ref{tab:sigmas} and summarized in
Sec.~\ref{sec:discussion}. The reader is urged to look first at the
table and discussion section and then to decide whether to read the
more detailed discussion in the main body of the text.

Section~\ref{sec:data} summarizes in a convenient form the data on the
measured solar neutrino event rates and Sec.~\ref{sec:calculations}
describes the way the calculations were done. Using the data and
methods described in the previous sections, Sec.~\ref{sec:results}
presents the principal results for the pre-SNO era, for just the water
Cherenkov solar neutrino detectors, for all six detectors, and for the
effect of taking account of the scaling of the fusion reactions that
produce $^7$Be and $^8$B solar neutrinos.
Section~\ref{sec:whymissing} explains physically why the $\chi^2_{\rm
min}$ solutions elliminate all $^7$Be and CNO neutrinos. I summarize
the main results briefly in Sec.~\ref{sec:discussion}.

All of the results presented in this paper depend upon the validity of
the published solar neutrino measurements. The formal statistical
estimates given here are not to be taken literally when the
probabilities become extremely small because of the likely effects of
unknown systematic errors.

\section{The measured rates}
\label{sec:data}

Table~\ref{tab:rates} summarizes the data from measurements of solar
neutrino event rates. For each of the six experiments listed in the
first column, the table gives in the second column the combined
prediction~\cite{bp2000} of the standard solar model and the simplest
version of standard electroweak theory (BP2000 solar model,no $\nu$
oscillations). The third column shows the measured values for each
experiment~\cite{sno,superk,chlorine,kamiokande,sage,gallexgno}. In
the last column, the table gives the ratio of each measured value to
the predicted standard value (no theoretical errors included in the
last column).  The dimensionless ratios are convenient to use in
calculations.  The last two rows present the weighted average rate for
the two $\nu-e$ scattering experiments (K and SK) and the weighted
average rate for the two gallium experiments (GALLIUM).
In this paper, we shall use the theoretical predictions only for one
very special case, when we compare the standard solar model with all
the measurements.

\begin{table}[!t]
\centering
\tightenlines
\caption[]{\baselineskip=12pt Solar Neutrino Rates: Standard theory
versus Experiment.  The unit is SNU ($10^{-36}$ interactions per
target atom per sec) for the radiochemical experiments:
Chlorine~\cite{chlorine}, SAGE~\cite{sage}, and GALLEX +
GNO~\cite{gallexgno}.  The unit is $10^6~{\rm cm^{-2}~s^{-1}}$ for the
water Cherenkov experiments, SNO~\cite{sno},
SuperKamiokande~\cite{superk}, and 
Kamiokande~\cite{kamiokande}, which measure the
${\rm ^8B}$ neutrino flux.  Results are also shown in the last two
rows for the weighted average of the SAGE and GALLEX/GNO experiments
and for the weighted average of the Kamiokande and Super-Kamiokande
experiments.  The BP2000 predictions for the combined standard solar
and electroweak model are taken from Ref.~\cite{bp2000}.  The errors
quoted for Measured/BP2000 are the quadratically combined statistical
and systematic uncertainties. The larger experimental error was used
here when asymmetric errors were quoted in the original publications.
\label{tab:rates}}
\begin{tabular}{lrrr}
\multicolumn{1}{c}{Experiment}&\multicolumn{1}{c}{BP2000}
&\multicolumn{1}{c}{Measured}&\multicolumn{1}{c}{Measured/BP2000}\\
\tableline
Chlorine&$7.6\left[1.00~^{+0.17}_{-0.14}\right]$&$2.56\left[1.00 \pm
0.088\right]$&$0.337 \pm 0.030$\\ 
Kamiokande&$5.05\left[1.00~^{+0.20}_{-0.16}\right]$&$2.80\left[1.00 \pm 0.136\right]$&$0.554 \pm
0.075$\\ 
SAGE&$128\left[1.00~^{+0.07}_{-0.05}\right]$&$77.0\left[1.00 \pm 0.087\right]$&$0.602 \pm
0.052$\\ 
GALLEX + GNO&$128\left[1.00~^{+0.07}_{-0.05}\right]$&$74.1\left[1.00 \pm 0.092\right]$&$0.579 \pm
0.053$\\ 
Super-Kamiokande&$5.05 \left[1.00~^{+0.20}_{-0.16}\right]$&$2.32 \left[1.00 \pm 0.037\right]$&$0.459 \pm
0.017$\\ 
SNO&$5.05\left[1.00~^{+0.20}_{-0.16}\right]$&$1.75\left[1.00
\pm 0.084\right]$&$0.3465 \pm 0.029$ \\
\tableline
\noalign{\smallskip}
K + SK&$5.05\left[1.00~^{+0.20}_{-0.16}\right]$&$2.34\left[1.00 \pm 0.035\right]$&$0.464 \pm
0.016$\\
GALLIUM&$128\left[1.00~^{+0.07}_{-0.05}\right]$&$75.6\left[1.00 \pm 0.063\right]$&$0.590 \pm
0.037$  
\end{tabular}
\end{table}

\section{Calculations}
\label{sec:calculations}

The predicted event rates are linear functions of the seven important
neutrino fluxes: $p$-$p$, $pep$, $hep$, ${\rm ^7Be}$, ${\rm ^8B}$, ${\rm
^{13}N}$, and ${\rm ^{15}O}$.
From purely physics considerations, any or all of these fluxes could be
important. In fact, at one time or another in the history of solar
neutrino research, each of these fluxes has been hypothesized to be
important for solar neutrino measurements \cite{book}.

\subsection{Equations and uncertainties}
\label{subsec:equations}

The equations for the neutrino event rates can be written conveniently in
terms of the ratios of the actual fluxes to the predicted BP2000
fluxes. In this case, the linear coefficients of the predicted solar neutrino
interaction rates can be read directly from Table~7 of
Ref.~\cite{bp2000} and the observed rates can be taken from the last
column of Table~\ref{tab:rates} of the present paper.  

The luminosity constraint can be written also as a convenient linear
equation in the neutrino fluxes. One has

\begin{equation}
1 = \sum_i \left({\alpha_i \over 10~{\rm MeV}}\right)
\left({\phi_i\over 8.532 \times 10^{10}~{\rm cm^{-2}s^{-1}}}\right),
\label{eq:lumconstraint}
\end{equation}
where accurate values of the energy coefficients $\alpha_i$ are given
in Ref.~\cite{luminositypaper} and the $\phi_i$ are the individual
neutrino fluxes ($i =  pp, pep, hep, {\rm ^7Be, ^8B, ^{13}N}$, and
${\rm ^{15}O}$).

The best-fit neutrino fluxes were obtained by minimizing $\chi^2$ for
each case considered (see Table~\ref{tab:sigmas} in the following
section for a description of the different cases). The $\chi^2$ can be
written symbolically as 

\begin{equation}
\chi^2 ~=~ \sum_i\frac{({\rm Rate}_i -\sum_j c_{ij} \phi_j)^2} 
{\sigma^2_{\rm exp} + \sigma^2_{\rm c.s.}} \, ,
\label{eq:chi2defn}
\end{equation}
where the $c_{ij}$ are numerical coefficients for each experiment
(cf. Table~7 of Ref.~\cite{bp2000}) and $\phi_j$ are the neutrino
fluxes.  The experimental errors, $\sigma_{\rm exp}$, were taken from
the last column of Table~\ref{tab:rates}. It is important to include
also the theoretical uncertainties for the calculated interaction
cross sections, $\sigma_{\rm c.s.}$. For each neutrino flux, the cross
section errors are take from Ref.~\cite{bahcall97} for the gallium
experiments and from Ref.~\cite{chlorcross} for the chlorine
experiment. The cross section uncertainties are included in the
reported rates for the other experiments listed in
Table~\ref{tab:rates}. Since the neutrino fluxes are treated as free
parameters, the uncertainties in the predicted fluxes are not included
in the calculations (except for the special case of testing the
standard solar model fit, cf. the last two rows of
Table~\ref{tab:sigmas}).

For a given number of degrees of freedom, $n$[$n =$ (number of
experiments $ + 1$) $-$ (number of free fluxes)] , the value of
$\chi^2_{\rm min}$ corresponds to a probability, $P$, that a worse fit
would have been obtained by chance if the model being tested is
correct. In our case, the model is that the measured experiments plus
the luminosity constraint are described by a theory in which the
undistorted neutrino energy spectra can have arbitrary
amplitudes. When the number of experiments plus the luminosity
condition is one more than the number of free-parameter neutrino
fluxes, then there is a particularly simple relation between
$\chi^2_{\rm min}$ and the effective number of $\sigma$'s. For this
special case ($n = 1$), $\sigma = \chi_{\rm min}$.  Here $\sigma$ is
the number of sigmas for a normal distribution such that the two-sided
probability of getting a value greater than $\sigma$ is equal to
$P$. More generally, for $\chi^2_{\rm min} \gg n$, one can
show\footnote{This asymptotic formula can be derived by integrating
the normal distribution by parts to obtain the leading term for the
probability to have a value greater than $\sigma$ and then equating
this expression to the leading term in the repeated fraction expansion
of the incomplete Gamma function that describes~\cite{pressetal} the
$\chi^2$ probability distribution.} that

\begin{equation}
\sigma^2 ~=~ \chi^2_{\rm min} - \ln\sigma^2 + (2 - n)\ln\chi^2_{\rm
min} +\ln g(n), 
\label{eq:sigmachi}
\end{equation}
where $g(n)= \Gamma^2(n/2)/(2^{1-n}\pi$) and $\Gamma$ is the Gamma
(generalized factorial) function. The result given in
Eq.~(\ref{eq:sigmachi}) is exact (not just asymptotically correct) for
$n$ equal to one.  For practical cases with $n \not= 1$,
Eq.~(\ref{eq:sigmachi}) can be solved simply by iteration with a hand
calculator.

\subsection{Supplementary conditions}
\label{subsec:supplementary}

We fit the results for at most six experiments and the luminosity
constraint. Therefore, we cannot use all seven of the neutrino fluxes
as free parameters. In previous free-flux analyses of solar neutrino
rates, nearly all authors have followed Hata {\it et
al.}~\cite{hata94} in taking the ratio of the $pep$ to $p$-$p$ fluxes
to be the same as in the standard solar model. The justification for
this assumption is that the $pep$ to $p$-$p$ ratio is practically
independent of details of the solar model, depending upon just the
weighted average of the density over the square root of the
temperature~\cite{bahcallmay69}.  For twelve variant and deviant solar
models listed in Table~10 of Ref.~\cite{bp2000}, the ratio of $pep$ to
$p$-$p$ fluxes is $(2.25 \pm 0.1)\times 10^{-3}$. The model that gives
the most extreme ratio (and also the most conservative result) is
ruled out by helioseismological data, giving a rms fit to the
helioseismological data that is more than $100$ times worse than the
standard solar model.  In the calculations described in the following
section, I chose the value of the $pep$ fluxes to be within the range
$(2.25 \pm 0.1)\times 10^{-3}$ $p$-$p$, varying the exact value to
give the most conservative result.

Many authors \cite{hata94,heeger96,bahcallkrastev96,bks98} have also
assumed that the CNO nuclear reactions are in equilibrium and have
therefore taken the $^{13}$N and $^{15}$O neutrino fluxes to be
exactly equal (or in some cases both to be equal to zero
\cite{parke95}).  Any non-zero value for the CNO neutrino fluxes
increases the discrepancy with the standard electroweak model.

Finally, all previous authors (except for Hata {\it et al.}
\cite{hata94}) have neglected the $hep$ flux, although $hep$ neutrinos
could in principle contribute significantly to the chlorine and
gallium experiments (see Ref. \cite{bp2000}).  At the $3\sigma$ upper
limit corresponding to the Super-Kamiokande result~\cite{superk}, the
$hep$ flux contributes $1.4\sigma_{\exp}{\rm (Cl)}$ (0.31 SNU) to
the chlorine rate but only $0.2 \sigma_{\exp}{\rm (Ga)}$ (0.8 SNU)
to the gallium rate. Here $\sigma_{\exp}{\rm (Cl)}$ is the total
experimental error for the chlorine experiment~\cite{chlorine} and $ 
\sigma_{\exp}{\rm (Ga)}$ is the total weighted average of the SAGE
and the GALLEX/GNO experiments.

I want to add a word of reassurance for the mathematically squeamish
who may be concerned about the fact that before imposing the
supplementary conditions there are more free fluxes than experiments
plus constraints. One can find the minimum $\chi^2$ using all the
fluxes. As we shall see in Section~\ref{sec:results} and
Section~\ref{sec:whymissing}, this minimum always lies in the region
within which the supplementary conditions apply, i. e., there are no
$^7$Be or CNO neutrinos. By choosing to consider the subset of fluxes
that give the smallest $\chi^2$, we are making it as difficult as
possible to reject the no-oscillation hypothesis.

In the following section, I explore the robustness of the free-flux
analyses to the supplementary conditions on $hep$ and CNO neutrinos described
above. For simplicity, I shall denote in Table ~\ref{tab:sigmas} and
in Sec.~\ref{sec:results} these conditions symbolically as:  i)
n13 = n15; and ii) $hep = 0.0$.

\section{Results}
\label{sec:results}

Table~\ref{tab:sigmas} presents the principal results of this paper.
The table gives the formal probability, P, of obtaining a fit as bad
($\chi^2 \geq \chi^2_{\rm min} $) as the best-obtainable fit with
arbitrary amplitudes, but undistorted energy spectra, for the solar
neutrino fluxes.  The table also gives the effective number of
standard deviations, $\sigma$ [defined by Eq.~(\ref{eq:sigmachi})], by
which the no neutrino oscillation hypothesis fails to fit the observed
data on solar neutrino event rates.  In a number of cases, the
probabilities quoted are so small that the distributions from which
the probabilities are calculated are not valid in the relevant extreme
limits. Therefore, I have included the effective number of sigmas
because most physicists have, based upon bitter experience with
unknown systematic errors, developed their own healthy internal
recalibration for the meaning of sigmas.

I also give in Table~\ref{tab:sigmas} the best-fit values, in units of
the BP2000 standard solar model fluxes~\cite{bp2000}, for the three
most important neutrino fluxes, $p$-$p$, $^7$Be, and $^8$B.  Contrary
to what one expects on astrophysical grounds, the formal minimization
process requires in all cases, where they are allowed to vary freely,
that the neutrino fluxes from the CNO chain, $^{13}$N and $^{15}$O, as
well as the $hep$ neutrino flux, be identically zero (for an
explanation, see Sec. \ref{sec:whymissing}). Therefore, these
fluxes are not given explicitly in Table~\ref{tab:sigmas}. The $pep$
neutrino flux is fixed to have that ratio relative to the basic
$p$-$p$ neutrino flux which produces the most conservative result,
given the general form of the ratio that results from weak interaction
theory (see discussion in Sec.~\ref{subsec:supplementary}).

The table also presents the predictions of each of the best-fits for
the capture rates in the radiochemical chlorine and gallium
experiments (in SNU, $10^{-36}$ interactions per target particle per
sec). By hypothesis, we are considering only undistorted energy
spectra; all the neutrinos are $\nu_e$. Therefore, the predicted rates
for the Kamiokande, Super-Kamiokande, and SNO CC measurements are,
when expressed in units of the predicted rates of the combined
standard model, all numerically equal to the tabulated value for the
$^8$B neutrino flux (which is given in units of the BP2000 flux).

One can make contradictory plausibility arguments about whether or not
one should use the weighted average of the
Kamiokande~\cite{kamiokande} and Super-Kamiokande~\cite{superk}
experiments or the weighted average of the SAGE~\cite{sage} and
GALLEX/GNO~\cite{gallexgno} experiments.  I therefore report
calculations performed with the experiments combined in different
ways. Fortunately, it turns out not to matter much whether one uses
the weighted averages or the individual experiments (see the section
labeled `Six experiments in different combinations' of
Table~\ref{tab:sigmas}). Because it yields the most conservative
answer, I adopt as `standard' the case in which one combines both
Kamiokande and Super-Kamiokande and SAGE and GALLEX/GNO\footnote{My
personal preference, however, is to regard the Super-Kamiokande and
Kamiokande measurements as two experiments because they have different
energy thresholds and because their energy calibrations were performed
in different ways. On the other hand, I prefer to combine the SAGE and
GALLEX/GNO experiments because they have exactly the same energy
sensitivity. One could argue, however, that for purposes of testing
the null hypothesis of no new physics the SAGE and GALLEX/GNO
experiments should be treated as different because they are located at
different places on earth and they made measurements over different
times.  The null hypothesis could, in principle, be wrong because of a
strong regeneration effect in the earth or because of a highly time
dependent neutrino flux.}.

\begin{table}[!t]
\tightenlines
\caption[]{How Many $\sigma$'s? The table shows the effective number
of standard deviations, $\sigma$, by which the no-oscillation
hypothesis fails to account for the total event rates measured in
solar neutrino experiments. The abbreviated notation for the six
experiments is the same as in Table~\ref{tab:rates}.  For each case
(combination of experiments), the table also gives the probability P
for errors distributed normally of obtaining a fit as bad as the
best-fit found, the neutrino fluxes (in units of the BP2000 fluxes)
for the $p$-$p$, $^7$Be, and $^8$B neutrinos, and the predicted event
rate (in SNU) for the chlorine and gallium experiments. The
supplementary conditions, $pep = pp$, $hep = 0.0$, n13 = o15, and n13
= o15 = 0.0 are defined in
Sec.~\ref{subsec:supplementary}. \label{tab:sigmas}}
\begin{tabular}{lccccccc}
Case&P&$\sigma$s&$pp$&${\rm
^7Be}$&${\rm ^8B}$&Cl&Ga\\
\hline
\noalign{\medskip}
\multicolumn{8}{c}{Pre-SNO}\\
\hline
\noalign{\smallskip}
Cl,K,Gallium, SK\tablenotemark[1]
&$1\cdot 10^{-2}$&2.5&1.0917&0.000&0.4550&2.9&84.6\\
\hline
\multicolumn{8}{c}{Only water Cherenkov experiments}\\
\hline
\noalign{\smallskip}
SK and SNO&$8 \cdot 10^{-4}$&3.35&---&---&0.4311&---&---\\
K, SK, and SNO&$1 \cdot 10^{-3}$&3.3&---&---&0.4356&---&---\\
\hline
\multicolumn{8}{c}{Six experiments in different combinations}\\
\hline
\noalign{\smallskip}
Cl,Gallium,K+SK,SNO\tablenotemark[1]&$7 \cdot 10^{-5}$&4.0&1.0917&0.000&0.4333&2.7&84.4\\
Cl,K,Gallium,SK,SNO\tablenotemark[2]&$3.5 \cdot 10^{-5}$&4.1&1.0917&0.000&0.4315&2.7&84.3\\
Cl,K,Gallium,SK,SNO\tablenotemark[3]&$3.5 \cdot 10^{-5}$&4.1&1.0917&0.000&0.4315&2.7&84.3\\
6 experiments&$3 \cdot 10^{-5}$&4.2&1.0917&0.000&0.4314&2.7&84.3\\
\noalign{\smallskip}
\hline
\multicolumn{8}{c}{${\rm ^7Be} \geq {\rm ^8B}$ in units of BP2000 fluxes}\\
\hline
\noalign{\smallskip}
Cl,Gallium,K+SK,SNO\tablenotemark[1]&$1 \cdot 10^{-13}$&7.4&1.039&0.6787&0.4144&3.4&103.6\\
Pre-SNO:~Cl,Gallium,K+SK\tablenotemark[4]&$4 \cdot 10^{-12}$&6.9&1.039&0.6907&0.4312&3.5&104.1\\
\noalign{\smallskip}
\hline
\multicolumn{8}{c}{Standard solar model}\\
\hline
\noalign{\smallskip}
Cl,Gallium,K+SK,SNO&$3 \cdot 10^{-11}$&6.7&1.000&1.000&1.000&7.6&127.8
\end{tabular}
\tablenotetext[1]{$ hep = 0.0$; n13 = o15.}
\tablenotetext[2]{$ hep = 0.0$.}
\tablenotetext[3]{n13 = o15.}
\tablenotetext[4]{$ hep = 0.0$; $n13 = 0.0$; $o15= 0.0$.}
\end{table}

\subsection{Pre-SNO}
\label{subsec:presno}

The situation prior to the announcement of the SNO results is the
first case listed in Table~\ref{tab:sigmas}. Considering the five
pre-SNO experiments, but using the weighted average of the SAGE and
GALLEX/GNO experiments, the no oscillation hypothesis is rejected at
the effective $2.5\sigma$ level ($99$\% C.L.) . Even this
most-favorable solution requires that $^7$Be neutrinos be entirely
missing, a result which many authors have argued is not physically or
astrophysically reasonable~\cite{bethe90,hata94,parke95,heeger96,spiro90,rosen94,bks98}.  

Following essentially all previous work on this subject, the case
listed in Table~\ref{tab:sigmas} includes the supplementary conditions
$pep = p\hbox{-}p$ , $hep = 0.0$, and n13= o15 (see explanation of
this notation in Sec.~\ref{subsec:supplementary}).  Nearly
identical results are obtained if SAGE and GALLEX/GNO are treated as
separate experiments and either pair of supplementary conditions, $pep
= p\hbox{-}p$ , $hep = 0.0$, or $pep = p\hbox{-}p$ and $n13 = o15$, is
used\footnote{The case in which SAGE and GALLEX/GNO are treated as
independent and the supplementary condition $hep = 0.0$ and $n13 =
o15$ is imposed leads to a doubly unphysical result for the parameters
that correspond to the minimum $\chi^2$. Not only is the $^7$Be flux
required to be zero, but also the best-fit $pep$ neutrino flux is
identically zero(cf. the extreme allowed range for $pep$ given in
Sec.~\ref{subsec:supplementary}). For this very unphysical case,
the $P = 0.07$, $^8$B = 0.4608, and the predicted chlorine and gallium
rates are, respectively, $2.7$ SNU and $81.8$ SNU.}.

\subsection{Water Cherenkov experiments}
\label{subsec:water}

The published results of the SNO CC and Super-Kamiokande experiments
are inconsistent at the level of $3.35\sigma$, as shown in
Ref.~\cite{sno} and in Table~\ref{tab:sigmas}. One might hope that
this result would be strengthened by including the 
Kamiokande measurement. However, this is not the case. The discrepancy
that arises from assuming no neutrino oscillations is
essentially unchanged if all three experiments are included; in this
case, the fit is acceptable at $P = 1\times10^{-3}$, which corresponds
to $3.3\sigma$ [for $\nu = 2$, cf. Eq.~(\ref{eq:sigmachi})].

Several authors have shown~\cite{villante98,fogli0102,fogli0106} how
one can chose the energy thresholds for the Super-Kamiokande and SNO
experiments such that the response functions for the two experiments
are made approximately equal. The advantage of this method is that
some of the systematic errors are reduced, but there is some slight
loss of statistical power. Also, one must understand the details of
the Super-Kamiokande experiment well enough to reevaluate accurately
the rate at a different threshold than the published
value. Apparently, this has been done successfully.  In obtaining the
results given in Table~\ref{tab:sigmas}, I have simply used the rates
and energy thresholds published by the Super-Kamiokande~\cite{superk}
and SNO~\cite{sno} collaborations. The straightforward result given in
the present paper is in good agreement with the more sophisticated
analysis described in Refs.~\cite{fogli0106,giunti01}.

\subsection{Six experiments in different combinations}
\label{subsec:six}
Table~\ref{tab:sigmas} shows the results for a variety of different
ways of combining all six of the experimental results and of imposing
the supplementary conditions.  The formal statistical probabilities of
obtaining fits as bad as the best fits that were found range from $P =
3\times10^{-5}$ to $P = 7\times10^{-5}$, about an order of magnitude
worse than obtained with the water Cherenkov experiments alone. The
corresponding number of sigmas at which the best-fit is formally
rejected is $4.0\sigma$ to $4.2\sigma$. 
The SNO experiment contributes $56$\% of the total $\chi^2$, with the
other experiments contributing much less: K + SK ($23$\%), gallium
($18$\%), and chlorine ($3$\%). Even if one omits without
justification the chlorine experiment, the result is barely affected;
$P = 3\times10^{-5}$ ($3.9\sigma$).

The best-fit solutions all correspond, as in the pre-SNO case, to the
unphysical result with an identically zero $^7$Be neutrino flux.

\subsection{Temperature scaling of nuclear reactions}
\label{subsec:temperature}

\begin{figure}[!t]
\tightenlines
\centerline{\psfig{figure=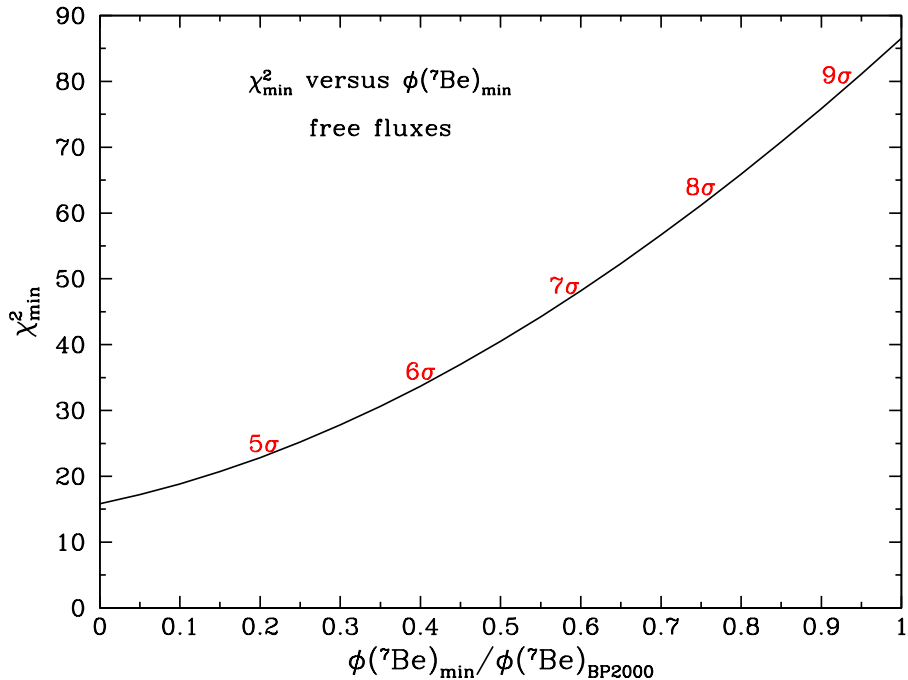,width=5in,angle=270}}
\caption[]{A little bit of $^7$Be goes a long way. 
The figure shows the best-fit $\chi^2_{\rm min}$ 
versus the minimum allowed value for $\phi ({\rm ^7Be})$. The neutrino
fluxes were treated as free parameters in fitting to the rates 
of the chlorine, gallium,  $\nu-e$ scattering, and SNO CC
experiments, except that  $\phi ({\rm ^7Be})$ was required to be at
least as large as $\phi ({\rm ^7Be})_{\rm min}$.  The neutrino flux is
measured in terms of the BP2000~\cite{bp2000} flux. Nuclear fusion
scaling arguments give $\phi ({\rm ^7Be})_{\rm min} \geq 0.69$. The
figure also shows the corresponding number of standard deviations at
which the no oscillation hypothesis is rejected.
\label{fig:be7}}
\end{figure}

The temperature scalings of the neutrino producing reactions can be
derived without considering details of a solar model, requiring only
energy-conservation and a quasi-static equilibriums~\cite{ulmer}. The
dominant factor is the exponential temperature dependence of the Gamow
penetration factor (see, e.g., Chapter 3 of Ref.~\cite{book}). In a
one zone model for the present-day sun, with a fixed temperature and
density, one finds~\cite{ulmer} for the $^7$Be neutrino flux
\begin{equation}
\phi(^7{\rm Be}) \propto T^{11} , 
\label{eq:be7temp}
\end{equation}

and for the $^8$B neutrino flux, 
\begin{equation}
\phi(^8{\rm B}) \propto T^{25} .
\label{eq:b8temp}
\end{equation}
These results are in excellent agreement with the scaling found in
detailed Monte Carlo studies of complete solar
models~\cite{bahcallulrich}. The scalings are robust because the Gamow
factor depends sensitively on temperature in the region of the
exponential tail where nuclear fusion reactions
occur~\cite{book,bethe,clayton,fowler}.

If the deficit of neutrinos observed in the water Cherenkov
experiments were due to astrophysical processes, then one would expect
that 
\begin{equation}
\left[\frac{\phi(^7{\rm Be})} {\phi(^7{\rm Be})_{\rm BP2000}}\right]
~\geq ~
\left[\frac{\phi(^8{\rm B})} {\phi(^8{\rm B})_{\rm
BP2000}}\right]^{11/25} \, .
\label{eq:be7min}
\end{equation}

Table~\ref{tab:sigmas} shows that, using all six experiments, the
best-fit for $ {\phi(^7{\rm Be})}/ {\phi(^7{\rm Be})_{\rm BP2000}} \,
\geq \, {\phi(^8{\rm B})} {\phi(^8{\rm B})_{\rm
BP2000}}$ is awful, corresponding to a rejection level,
$7.4\sigma$, that is so small that it has no practical meaning.
The gallium experiments (SAGE and GALLEX/GNO)  contribute $50$\% of 
the total $\chi^2$, with the chlorine and K + SK both contributing
$\sim 20$\%.

For practical purposes, the fit omitting the  SNO result is just about as
bad, $6.9\sigma$, as the fit including SNO.

Figure~\ref{fig:be7} shows the
dependence of $\chi^2_{\rm min}$ on the minimum allowed value of $
{\phi(^7{\rm Be})}$. It is obvious from Fig.~\ref{fig:be7} that a
little $^7$Be goes a long way.

\subsection{Standard solar model}
\label{subsec:standard}

The fit with the standard solar model, BP2000~\cite{bp2000}, is shown
in the last row of Table~\ref{tab:sigmas}. The fit is bad, but the
formal level of the discrepancy is not quite as bad as for the
free-parameter case with $ {\phi(^7{\rm Be})}/ {\phi(^7{\rm Be})_{\rm
BP2000}} \, \geq \, {\phi(^8{\rm B})} {\phi(^8{\rm B})_{\rm
BP2000}}$. The reason for the somewhat lower apparent discrepancy is
that for the standard solar model the relatively large theoretical
uncertainties in the flux predictions are included in the quadratically
added errors.

Nevertheless, the agreement between the standard model predictions and
the experimental measurements is poor, with formal values of $\sigma =
6.7$ and $P = 3\times 10^{-11}$.

\section{Why are the $^7$Be, $^{13}$N, $^{15}$O, and $hep$ neutrinos
required to be missing completely?}
\label{sec:whymissing}

All of the formal best-fits of the undistorted solar neutrino energy
spectra to the observed event rates require that the $^7$Be, $^{13}$N,
$^{15}$O, and $hep$ neutrino fluxes be identically zero, On the other
hand, these same solutions require a $^8$B flux that is a reasonable
compromise fit to the observed fluxes of $^8$B neutrinos in the
different water Cherenkov detectors (no big surprise) and a $p-p$
neutrino flux that is typically $9$\% larger than the predicted
standard solar model $p-p$ neutrino flux (cf. Table~\ref{tab:sigmas}).

The requirement that $^7$Be, $^{13}$N, $^{15}$O, and $hep$ neutrino
fluxes be absent contradicts all astrophysical calculations and basic
laboratory astrophysics
data~\cite{bethe90,hata94,parke95,heeger96,spiro90,rosen94,bks98}. For
example, the formal solutions require the complete absence of $^7$Be
neutrinos and only a modest reduction in the flux of $^8$B neutrinos
predicted by the standard solar model.  But, both $^7$Be and $^8$B
neutrinos are produced by nuclear interactions on the same parent
isotope ($^7$Be), with production of $^7$Be neutrinos being about a
thousand times more probable (according to the standard solar model
estimates). Moreover, according to simple estimates and to detailed
calculations, the fusion reactions that lead to $^{13}$N and $^{15}$O
neutrinos occur more frequently than the reaction which gives rise to
$^8$B neutrinos.

So why do the minimum $\chi^2$ solutions prefer $p-p$ neutrinos and
abhor $^7$Be, $^{13}$N, $^{15}$O, and $hep$ solar neutrinos? The
answer is simple and is contained in the basic equation that describes
the nuclear fusion process that is responsible for the solar
luminosity: four protons are burned to form an alpha-particle, two
positrons, and two electron-type neutrinos. Thus

\begin{equation}
4p \to \alpha + 2e^+ + 2\nu_e \, .
\label{eq:hburning}
\end{equation}
Equation~(\ref{eq:hburning}) shows that two neutrinos are emitted every
time four protons are burned to an alpha-particle.

The $\chi^2_{\min}$ solutions prefer replacing other neutrinos by
$p-p$ neutrinos since $p-p$ neutrinos have by far the lowest energies
($\leq 0.42 $ MeV). Because of their low energies, $p-p$ neutrinos
have the smallest interaction cross sections in gallium solar neutrino
detectors and are not detected at all in the chlorine and water
Cherenkov experiments.  Moreover, the amount of thermal energy
communicated to the star depends upon which neutrinos are emitted: if
high energy neutrinos are emitted, then less energy is communicated to
the star.

If one replaces in a formal fitting process a higher energy neutrino
(like $^7$Be or $^{13}$N ) by a $p-p$ neutrino, then mathematically
the replaced solution wins in two ways: more thermal energy is communicated to
the star (making it easier to satisfy the luminosity constraint) and
the calculated event rates are lower (in agreement with observations).

\section{Discussion}
\label{sec:discussion}

No satisfactory fit can be found that describes well the observed
event rates in solar neutrino experiments, if the different sources of
solar neutrinos have amplitudes that can be treated as free parameters
but energy spectra that are undistorted.

Table~\ref{tab:sigmas} summarizes quantitatively the situation
regarding fits with undistorted energy spectra to the measured solar
neutrino event rates. For the pre-SNO era, which includes five
experiments (chlorine, Kamiokande, SAGE, GALLEX, and
Super-Kamiokande), the no-distortion, no-oscillation hypothesis fails
at a formal statistical level of $99$\% ($2.5\sigma$). Considering
just the water Cherenkov experiments, SNO, Super-Kamiokande, and
Kamiokande, the discrepancy with the no-oscillation hypothesis is at
the $3.3\sigma$ level, where the formal probability of obtaining a fit
as bad as observed is $P = 8\times10^{-4}$. These results confirm
previously published calculations and are included here to provide an
appropriate context. 

Two new results on the quality of fits to solar neutrino data are
presented in this paper. First, if all six of the solar neutrino
experiments (SNO, chlorine, Kamiokande, SAGE, GALLEX, and
Super-Kamiokande) are included in the fit, the formal statistical
level at which the fit is unsatisfactory rises to $4.0\sigma$ ($P =
7\times10^{-5}$). Second, if the temperature scaling of the fusion
reactions producing the $^7$Be and $^8$B solar neutrinos are taken
into account, then the failure of the fitting procedure is at the
enormously high $7.4\sigma$ level (the formal value of $P$ is ridiculously
small).

To achieve even the unsatisfactory fits described in
Table~\ref{tab:sigmas}, the $\chi^2_{\rm min}$ solutions require that
the $^7$Be, $^{13}$N, and $^{15}$O neutrinos be entirely
missing. Section~\ref{sec:whymissing} describes the physical reasons
leading to this unphysical effect.  Since the formal best-fits are
achieved at the expense of eliminating the $^7$Be, $^{13}$N, and
$^{15}$O neutrinos, we should really regard the (improbable) solutions
found here as even more unlikely because of their physical implausibility.

\acknowledgments I am grateful to L. Bergstrom for a question that
sparked this investigation and to L. Bergstrom, A. Friedland, W. Haxton,
P. Langacker, and W. Press for valuable discussions and suggestions.  
I acknowledge valuable discussions and stimulating collaborations 
with P. Krastev in prior work on this subject.
This research is
supported by NSF Grant No. PHY0070928.


\begin{references}
\bibitem{sno}Q. R. Ahmad {\it et al.} (SNO collaboration),
Phys. Rev. Lett. {\bf 87}, 071301 (2001).
\bibitem{superk}S.~Fukuda {\it et al.}  (Super-Kamiokande Collaboration),{\it
Phys. Rev. Lett.} {\bf 86}, 5651 (2001).
\bibitem{bp2000}J. N. Bahcall, M. H. Pinsonneault, and S. Basu,
Astrophys. J. {\bf 555}, 990 (2001).
\bibitem{bahcall94}J. N. Bahcall, Phys. Rev. D {\bf 49}, 3923 (1994).
\bibitem{chlorine}B. T. Cleveland {\it et al.}, Astrophys. J. {\bf
496}, 505 (1998).
\bibitem{kamiokande}Y. Fukuda {\it et al.},  Phys. Rev. Lett. {\bf 77},
 1683 (1996).
\bibitem{sage}J. N. Abdurashitov {\it et al.} (SAGE Collaboration), 
 Phys. Rev. C {\bf 60}, 055801 (1999). The latest results, from
April 16, 2001, are given at the SAGE web
site:http://EWIServer.npl.washington.edu/SAGE/SAGE.html .
\bibitem{gallexgno}M. Altmann {\it et al.} (GNO Collaboration), 
Phys. Lett. B {\bf 490}, 16 (2000); W. Hampel {\it et al.} (GALLEX
Collaboration), Phys. Lett. B {\bf 447}, 127 (1999).
\bibitem{luminositypaper}J. N. Bahcall, Phys. Rev. D (submitted).
\bibitem{bethe90}J. N. Bahcall and H. A. Bethe, Phys. Rev. Lett. {\bf
65}, 2233 (1990).
\bibitem{hata94}N. Hata, S. Bludman, and P. Langacker, Phys. Rev. D
{\bf 49}, 3622 (1994).
\bibitem{parke95}S. Parke, Phys. Rev. Lett. {\bf 74}, 839 (1995).
\bibitem{heeger96}K. M. Heeger and R. G. H. Robertson,
Phys. Rev. Lett. {\bf 77}, 3720 (1996).
\bibitem{spiro90}M. Spiro and D. Vignaud, Phys. Lett. B {\bf 242}, 279
(1990); A. Dar and S. Nussinov, Particle World {\bf 2}, 117 (1991);
V. Castellani {\it et al.}, Phys. Lett. B {\bf 324}, 425 (1994); {\bf
329}, 525(E) (1994); V. Berezinsky, Comments Nucl. Part. Phys. {\bf
21}, 249 (1994); G. L. Fogli, E. Lisi, and D. Montanino, Phys. Rev. D
{\bf 49}, 3226 (1994); see also G. L. Fogli and E. Lisi,
Astropart. Phys. {\bf 13}, 185 (1995).
\bibitem{bahcallkrastev96}J. N. Bahcall and P. I. Krastev, Phys. Rev. D 
{\bf 53}, 4211 (1996).
\bibitem{bks98}J. N. Bahcall, P. I. Krastev, and A. Yu. Smirnov,
Phys. Rev. D {\bf 58}, 096016 (1998).
\bibitem{book}J. N. Bahcall, {\it Neutrino Astrophysics} (Cambridge
University Press, Cambridge, England, 1989).
\bibitem{bahcall97}J. N. Bahcall, Phys. Rev. C {\bf 56}, 3391 (1997).
\bibitem{chlorcross}J. N. Bahcall, L. Lisi, D. E. Alburger, L. 
De~Braeckeleer, S. J. Freedman, and J. Napolitano, Phys. Rev. C
{\bf 54}, 411 (1996).
\bibitem{pressetal}W. H. Press, B. P. Flannery, S. A. Teukolsky, and
W. T. Vettering, {\it Numerical Recipes} (Cambridge University Press,
Cambridge, England, 1986).
\bibitem{bahcallmay69}J. N. Bahcall and R. M. May, Astrophys. J. {\bf
155}, 501 (1969).
\bibitem{rosen94}W. Kwong and S. P. Rosen, Phys. Rev. Lett. {\bf 73},
369 (1994); J. N. Bahcall, Phys. Lett. B {\bf 338}, 276 (1994).
\bibitem{villante98}F. L. Villante, G. Fiorentini, and E. Lisi,
Phys. Rev. D {\bf 59}, 013006 (1999). 
\bibitem{fogli0102}G. L. Fogli, E. Lisi, A. Palazzo, and
F. L. Villante, Phys. Rev. D {\bf 63}, 113016 (2001).
\bibitem{fogli0106}G. L. Fogli, E. Lisi,  D. Montanino, and  A. Palazzo,
submitted for publication, hep-ph/0106247.  
\bibitem{giunti01}C. Giunti, submitted for publication, hep-ph/0107310.
\bibitem{ulmer}J. N. Bahcall and A. Ulmer, Phys. Rev. D {\bf 53}, 4202
(1996). 
\bibitem{bahcallulrich}J. N. Bahcall and R. K. Ulrich,
Rev. Mod. Phys.  {\bf 60}, 297 (1988).
\bibitem{bethe}H. A. Bethe, Phys. Rev. {\bf 55}, 434 (1939).
\bibitem{clayton}D. D. Clayton, {\it Principles of Stellar Evolution
and Nucleosynthesis} (University of Chicago Press, Chicago, 1983).
\bibitem{fowler}W. A. Fowler, Rev. Mod. Phys. {\bf 56}, 149 (1984).
\end{references}
\end{document}